\documentclass[preprint]{aastex}
\usepackage{lineno}
\usepackage{xcolor}

\begin{document}

\title {Absolute Properties of the Oscillating Eclipsing Algol X Trianguli }
\author{Jae Woo Lee$^{1}$, Kyeongsoo Hong$^{2}$, Jang-Ho Park$^{1}$, Marek Wolf$^{3}$, and Dong-Jin Kim$^{1}$ }
\affil{$^1$Korea Astronomy and Space Science Institute, Daejeon 34055, Republic of Korea}
\affil{$^2$Institute for Astrophysics, Chungbuk National University, Cheongju 28644, Republic of Korea}
\affil{$^3$Astronomical Institute, Faculty of Mathematics and Physics, Charles University in Prague, 180 00 Praha 8, V Hole\v sovi\v ck\'ach 2, Czech Republic}
\email{jwlee@kasi.re.kr}

\begin{abstract}
We report results from the TESS photometric data and new high-resolution spectra of the Algol system X Tri showing short-period 
pulsations. From the echelle spectra, the radial velocities of the eclipsing pair were measured, and the rotational rate and 
effective temperature of the primary star were obtained to be $v_1$$\sin$$i=84\pm6$ km s$^{-1}$ and $T_{\rm eff,1}=7900 \pm 110$ K, 
respectively. The synthetic modeling of these observations implies that X Tri is in synchronous rotation and is physically linked 
to a visual companion TIC 28391715 at a separation of about 6.5 arcsec. The absolute parameters of our target star were accurately 
and directly determined to be $M_1 = 2.137 \pm 0.018$ $M_\odot$, $M_2 = 1.101 \pm 0.010$ $M_\odot$, $R_1 = 1.664 \pm 0.010$ $R_\odot$, 
$R_2 = 1.972 \pm 0.010$ $R_\odot$, $L_1 = 9.67 \pm 0.55$ $L_\odot$, and $L_2 = 2.16 \pm 0.09$ $L_\odot$. 
The phase-binned mean light curve was used to remove the binary effect from the observed TESS data. Multifrequency analysis of 
the residuals revealed 16 significant frequencies, of which the high-frequency signals between 37 day$^{-1}$ and 48 day$^{-1}$ 
can be considered probable pulsation modes. Their oscillation periods of 0.021$-$0.027 days and pulsation constants of 
0.014$-$0.018 days are typical values of $\delta$ Sct variables. The overall results demonstrate that X Tri is an oEA star system, 
consisting of a $\delta$ Sct primary and its lobe-filling companion in the semi-detached configuration. 
\end{abstract}


\section{INTRODUCTION}

Most stars in binary and multiple systems are born almost simultaneously and evolved by interaction between their components 
(Duch\^ene \& Kraus 2013). The pulsations in such systems are very useful for probing the impacts of mass transfer and tides on 
stellar evolution and interior structure (Murphy 2018; Bowman et al. 2019). Many of them are $\delta$ Sct pulsations in 
the mass-accreting primary components of semi-detached interacting Algols (Zhou 2010; Lee et al. 2016; 
Kahraman Ali\c cavu\c s et al. 2017; Liakos \& Niarchos 2017; Mkrtichian et al. 2018), the so-called oscillating eclipsing Algol 
(oEA) class (Mkrtichian et al. 2002, 2004). In oEA precursors, the original more massive components fill their Roche lobes and 
some masses are transferred to the less massive companions. As a consequence of the mass transfer and accretion, the gainers become 
the present A/F-type pulsating primary components and the losers become the current low-mass secondary stars of spectral type F$-$K 
that are magnetically active. 

For eclipsing $\delta$ Sct stars, a possible relationship between the orbital and pulsation periods was established empirically and 
theoretically, respectively, by Soydugan et al. (2006) and Zhang et al. (2013); the longer the former ($P_{\rm orb}$), the longer 
the latter ($P_{\rm pul}$). Liakos \& Niarchos (2015, 2017) suggested that the oscillations are influenced by the binarity in 
$P_{\rm orb}<$ 13 days. However, Kahraman Ali\c cavu\c s et al. (2017) showed that the threshold is too low and the $\delta$ Sct 
pulsators in eclipsing binaries (EBs) oscillate in shorter pulsation periods with lower amplitudes than single variables. 
Mkrtichian et al. (2018) reported that the rapid mass accretion on the oscillating primary in the oEA system RZ Cas is produced 
by the magnetic cycle of its lobe-filling companion, and this can induce a periodic variation in pulsation properties. 
Recently, many new $\delta$ Sct-like pulsators in EBs have been detected from the archival data of the TESS mission, but 
their physical properties remain unclear (e.g., Kahraman Ali\c cavu\c s et al. 2022; Shi et al. 2022). A review of such pulsating 
EBs was presented in Lampens (2021). 

We have been carrying out follow-up spectroscopy of the EB systems with pulsating components. The Algol system X Tri (TIC 28391714; 
Gaia EDR3 298268990328628224; TYC 1763-2733-1; $T_{\rm p}$ = $+$8.565; $V\rm_T$ = $+$9.036, $(B-V)\rm_T$ = $+$0.328; $P_{\rm orb}$ 
= 0.9715 days) has been included as a probable oscillating EB in our target list. It was individually discovered by Walker (1921) 
and Neujmin (1922) as a variable star. The previous spectra of the program target were secured by Struve (1946). He classified 
the spectral types of both components as A3$+$G3 and obtained only primary radial velocities (RVs) with a semi-amplitude of 
$K_1$ = 110 km s$^{-1}$ and a rotation rate of $v_1$ = 50 km s$^{-1}$, about 0.56 times slower than its synchronous value 
(van Hamme \& Wilson 1990). Bozkurt et al. (1976) and Liakos et al. (2010) made complete photoelectric and CCD light curves in 
the $BV$ and $BVRI$ bands, respectively, and analyzed their own observations. Mezzetti et al. (1980) solved again the multiband 
light curves of Bozkurt et al. (1976) with the elements of spectroscopic orbits of Struve (1946) and presented the masses and 
radii of each component: $M_1$ = 2.3 $\pm$ 0.7 $M_\odot$, $M_2$ = 1.2 $\pm$ 0.3 $M_\odot$, $R_1$ = 1.71 $\pm$ 0.03 $R_\odot$, and 
$R_2$ = 1.96 $\pm$ 0.03 $R_\odot$. The historical results indicate that the program target has a semi-detached configuration with 
the lobe-filling secondary. Meanwhile, its orbital period has been studied by Gadomski (1932), Wood (1950), Wood \& Forbes (1963), 
Frieboes-Conde \& Herczeg (1973), Rafert (1982), Rovithis-Livaniou et al. (2000), and Qian (2002). Most recently, Liakos et al. (2010) 
suggested that X Tri is likely a quintuple system with three circumbinary companions in periods and minimal masses of 36.9 yrs and 
0.18 M$_\odot$, 22.4 yrs and 0.24 M$_\odot$, and 16.8 yrs and 0.22 M$_\odot$.

There have been some photometric attempts to search for pulsation features in X Tri. Kim et al. (2003) and Liakos \& Niarchos (2009) 
did not find any noticeable signals from their 2001 and 2007$-$2008 observations, respectively, while Turner et al. (2014) detected 
an oscillation with a period of 0.0220 days and an amplitude of 20 mmag in $V$ passband during their campaign on October 2010. 
This article is the tenth in a series studying the absolute properties of the pulsating EBs by combining our high-resolution spectra 
with existing or new photometric data (cf. Hong et al. 2015; Kim et al. 2022).

\section{TESS PHOTOMETRY AND ECLIPSE TIMES}

X Tri was observed in Sector 17 (BJD 2,458,764.688 $-$ 2,458,789.684) by camera 1 of TESS (Ricker et al. 2015) and recorded 
with a 2-min cadence. The observations had an interruption of about 1.5 days near the midpoint for data download. For this study, 
we used the \texttt{SAP$_-$FLUX} data available from the MAST portal\footnote{https://archive.stsci.edu/} and did not include 
all points between BJD 2,458,772.466 and 2,458,776.278 showing unrealistic light variations. The raw data were detrended and 
normalized by applying a second-order polynomial fit to each of the two segments divided by the interruption (Lee et al. 2019). 
We transformed the corrected fluxes to magnitudes. 

From the time-series data, 38 mid-eclipse timings were calculated with the Kwee \& van Woerden (1956) method and are compiled in 
Table 1. With the timing measurements, we derived the linear eclipse ephemeris for the TESS observations, as follows: 
\begin{equation}
 \mbox{Min I} = \mbox{BJD}~ 2,458,765.148638(5) + 0.9715257(3)E. 
\end{equation}
The orbital period ($P_{\rm orb}$) is equivalent to a frequency of $f_{\rm orb}$ = 1.0293088 $\pm$ 0.0000005 day$^{-1}$. Using 
this ephemeris, the phase-folded light curve of X Tri is presented in Figure 1. 

Using all available minimum epochs, we plotted the current eclipse timing $O-C$ diagram of X Tri, which is given in 
the upper panel of Figure 2. In this study we collected 653 times of minimum light. Most of the eclipse times can be found f.e. 
in the $O-C$ Gateway database\footnote{http://var2.astro.cz/ocgate/index.php?lang=en}. The general trend of the $O-C$ values is 
a downward parabola, suggesting a long-term period decrease in this system. With the least-squares method, we derived 
the following quadratic ephemeris: 
\begin{equation}
 C = \mbox{HJD}~ 2,447,296.2617(3) + 0.97152997(5)E - 1.91(2) \times 10^{-10}E^2. 
\end{equation}
As was claimed by previous investigators, the secular period decrease cannot be connected with a mass transfer in the system. 
In case of a conservative mass transfer in such a semidetached system with the Roche-lobe filling secondary and more massive primary 
component, the orbital period should be increasing. The timing residuals from the quadratic ephemeris are presented in the lower 
panel. One can see irregular or quasi-periodic oscillations with typical lengths of about 20 years, which cannot be explained simply 
by a light-time effect (LITE) due to a circumbinary object. In this context, Liakos et al. (2010) suggested a multiple LITE of 
three additional bodies, which partially explain these variations. 

X Tri is a member of a similar short-period Algol group (TU Her, FH Ori, TY Peg, Z Per, Y Psc) with a known long-term period 
decrease as well as short-period changes (Qian 2002). The nature of the long-term decrease as well as short-term irregular period 
changes has not yet been clearly and convincingly explained, and are often attributed to the mass and angular momentum loss 
mechanism, as developed by Tout \& Hall (1991). Concerning a possible magnetic activity cycle in the secondary companion, there is 
not enough energy in the outer layers of the active cool star to reproduce the observed orbital period changes (Rovithis-Livaniou et al. 2000).

\section{NEW SPECTROSCOPY AND DATA ANALYSIS}

We conducted time-series spectroscopy through the BOES spectrograph mounted on the 1.8 m telescope at the Bohyunsan Optical 
Astronomy Observatory (Kim et al. 2007). Thirty-five spectra of X Tri were acquired on five nights between December 2019 and 
November 2020. Their wavelength range and resolving power are $\lambda$ = 3600$-$10,200 $\rm \AA$ and $R$ = 30,000, respectively. 
The integration time of the program target was set to 30 min, corresponding to 0.021 of the eclipse period. The raw CCD images 
were processed utilizing the IRAF packages CCDRED and ECHELLE. The reduction process was identical to that employed by 
Hong et al. (2015). The consequential signal-to-noise ratio (SNR) is approximately 40, measured between 4000 $\rm \AA$ and 5000 $\rm \AA$. 

Struve (1946) reported that his results were not fully exempted from rotational broadening and line blending between the components. 
Trailed spectra are of great use to reveal and track the orbital motions of binary stars (e.g., Lee et al. 2018, 2020a,b). In order 
to look for sets of absorption lines for X Tri, we made the phase-folded trailed spectra from the BOES observations and examined them 
in detail at full wavelength. As a consequence, we found that \ion{Fe}{1} $\lambda$4957.61 and \ion{Ca}{1} doublet $\lambda$6103/6122 
lines are strong enough to individually trace both component stars. 

In Figure 3, the trailed spectra of the \ion{Ca}{1} doublet are displayed in the upper panel and the spectrum observed on 
BJD 2,459,169.0452 (0.73 phase) is presented as an example in the lower panel. For RV determinations, the absorption lines were 
fitted several times by means of double Gaussian functions with the deblending routine in the IRAF $splot$ task. The final RVs 
and their 1-$\sigma$ values were averaged and standardized from these measurements. They are presented in Table 2 and Figure 4, 
where triangles and circles denote \ion{Fe}{1} and \ion{Ca}{1} RVs, respectively. From a sine-curve fit to each star's RVs, 
we obtained the preliminary semi-amplitudes of $K_1$ = 107.8 km s$^{-1}$ and $K_2$ = 202.2 km s$^{-1}$, respectively, corresponding 
to a mass ratio of $q$ = 0.533. 

The atmospheric parameters of X Tri can be measured by comparing its observed spectra and synthetic models (Hong et al. 2017; 
Lee et al. 2022b). First of all, we constructed a disentangled spectrum from our BOES spectra and RV measures using the FDB\textsc{inary} code 
of Iliji\'c et al. (2004). Then, the projected rotational velocity ($v_1$$\sin$$i$) and effective temperature ($T_{\rm eff,1}$) 
of the primary component were yielded by minimizing the $\chi^{2}$ statistic between the reconstructed spectrum and model spectra. 
To do this, we generated grids of synthetic spectra in ranges of 10 $\le$ $v_1$$\sin$$i$ $\le$ 200 km s$^{-1}$ and 
6000 $\le$ $T_{\rm eff,1}$ $\le$ 10,000 K from the BOSZ spectral library (Bohlin et al. 2017). The surface gravity of $\log g_1$ 
= 4.3 (cf. Section 6) and the solar metallicity of $[$Fe/H$]=0$ were applied to the calculations of the model spectra, because 
the two parameters are highly correlated with stellar temperatures. For the $\chi^{2}$ minimization, we selected seven spectral 
regions (\ion{Ca}{2} $\lambda$3933, H$_{\rm \delta}$ $\lambda$4101, H$_{\rm \gamma}$ $\lambda$4340, \ion{Fe}{1} $\lambda$4383, 
H$_{\rm \beta}$ $\lambda$4861, \ion{Ca}{1} doublet $\lambda$6103/6122, H$_{\rm \alpha}$ $\lambda$6563), which are displayed in 
Figure 5. We acquired the best-fitting parameters of $v_1$$\sin$$i=84\pm6$ km s$^{-1}$ and $T_{\rm eff,1}=7900 \pm 110$ K. 
These values indicate that the primary star is rotating $\sim$1.7 times faster than Struve's (1946) rate of 50 km s$^{-1}$ and 
its spectral type is deduced to be A6.5 V (Pecaut \& Mamajek 2013). 

On the other hand, there is no evidence of any Balmer emission nor the presence of a circumstellar matter in our observed spectra. 
This may be because the emission feature is very weak and/or our SNR is not enough to detect it.

\section{BINARY MODELING AND ABSOLUTE DIMENSIONS}

Like the historical data (Bozkurt et al. 1976; Liakos et al. 2010), the TESS light curve of X Tri has similarities to that of 
classical Algols. To see if there was any light variability in X Tri, we divided the time-series TESS data into 19 light curves 
at intervals of $P_{\rm orb}$ = 0.9715257 days and measured each of them at four orbital phases 0.0 (Min I), 0.25 (Max I), 
0.5 (Min II), and 0.75 (Max II). These measurements appear in Table 3, where the BJD time of each segment is the middle of 
the beginning and the end. The average values at each phase are given in the last line of Table 3. As presented in this table, 
the primary eclipse depth is more than five times ($\sim$1.24 mag) deeper than the secondary one and the depth difference implies 
a large temperature difference between the component stars. Further, Max I is $\sim$0.006 mag brighter than Max II, which may be 
indicative of stellar activities on and around both components. The light levels for each light curve were subtracted from 
the TESS average for the whole datasets and the mean residuals are plotted in Figure 6. We show that the primary minima gradually 
brightened up after the interruption in the middle point. 

The TESS archive data and our velocity curves of X Tri were solved applying the Wilson-Devinney (W-D) code (Wilson \& Devinney 1971, 
van Hamme \& Wilson 2007) with the proximity and eclipse effects on RVs, as in the case of V404 Lyr (Lee et al. 2020a) and 
WASP 0131+28 (Lee et al. 2020b). The main difference between the 2007 version we used and the latest version is 
(1) direct distance estimation and (2) the use of absolute physical units (e.g., Kallrath, J. 2022), which does not affect 
our X Tri parameters. Thus, we used the 2007-version W-D code with features optimized by us, including repetitive routines. 
The absolute dimensions and distance determination of X Tri were calculated independently regardless of the W-D code. 

In our synthesis, we fixed the primary's temperature at $T_{\rm eff,1} = 7900 \pm 110$ K measured from 
the BOES spectra. It is assumed that X Tri A and B have predominately radiative and convective atmospheres, respectively. 
Bolometric albedo and gravity-darkening exponent were set to be ($A_1$, $g_1$) = (1.0, 1.0) and ($A_2$, $g_2$) = (0.5, 0.32). 
The logarithmic law was used for the limb-darkening effect (van Hamme 1993). Based on $v_1$$\sin$$i=84\pm6$ km s$^{-1}$ and 
for the lobe-filling secondary, we adopted the rotation parameters of $F_{1,2}$ = 1.0 in the W-D synthetic code.

In a simultaneous light and RV solution, it is hard to assign reasonable weights to these observations. To resolve this, the binary 
modeling in this article was conducted in two steps. To begin with, the TESS light curve was analyzed using the $q$ value taken 
from our RV fitting. In the second, our RV curves were fitted by adopting the light curve parameters derived in the first step. 
Then again, the TESS data were analyzed applying the newly calculated spectroscopic elements ($q$, $\gamma$, and $a$). This process 
was carried out for two cases without and with a starspot, and was repeated until both curves were reasonably solved. 

Table 4 presents the final model parameters with a cool starspot located on the secondary's surface, and the RV synthetic curves 
from them are shown as red lines in Figure 4. The red line in the top panel of Figure 1 presents our spot solution, 
and the light curve residuals corresponding to the unspotted and cool-spot models are given in the second and third panels. 
Although the light diminution seen between phases 0.5 and 0.8 was better described by the spot model, we could not fit 
the quasi-sinusoidal modulations to occur around both eclipses. The `W'-shape variation pattern of the residuals could result 
from some combination of stellar phenomena, such as magnetic activity, a gas stream from the lobe-filling component, and 
its impact on the detached companion. However, at present, we do not have a clear explanation for this puzzle. 
As in our previous studies (Lee et al. 2021, 2022a), the uncertainties of the binary star parameters in Table 4 were computed 
following the procedure applied by Southworth et al. (2020) for the TESS time-series data. 

The binary star model indicates that X Tri is a semi-detached Algol, in which the detached primary component fills about 69 \% of 
its inner Roche lobe. The absolute dimensions of each star in Table 5 were computed from the synthetic solution of the TESS data and
our double-lined RVs. We used the solar magnitude of $M_{\rm bol}$$_\odot$ = +4.73 and the empirical bolometric corrections (BCs) 
presented by Torres (2010). 
For measuring the distance to X Tri, we first obtained its apparent magnitude and color index (CI) to be $V$ = +9.01 $\pm$ 0.02 
and ($B-V$) = 0.28 $\pm$ 0.03 mag, respectively, from the Tycho magnitudes of $B\rm_T$ and $V\rm_T$ (H\o g et al. 2000). Then, 
the intrinsic CI of ($B-V$)$\rm_0$ = +0.18 $\pm$ 0.02 was estimated applying $T_{\rm eff,1}$ to the temperature$-$color calibration 
(Flower 1996), and $E$($B-V$) = +0.10 $\pm$ 0.04. Our computed values lead to a nominal distance of 215 $\pm$ 6 pc, which is in 
good agreement with 210 $\pm$ 2 pc inverted from GAIA EDR3 ($\pi$ = 4.753 $\pm$ 0.042 mas; Gaia Collaboration et al. 2021).

\section{PULSATIONAL CHARACTERISTICS}

The fundamental parameters of X Tri in Table 5 indicate that the hotter primary star is present in the main-sequence (MS) $\delta$ Sct 
region of the Hertzsprung-Russell diagram, while the less massive secondary lies above the MS band where the lobe-filling companions 
of other semi-detached Algols exist (\. Ibano\v{g}lu et al. 2006; Lee et al. 2016). 
Turner et al. (2014) found a pulsation period of 0.0220 $\pm$0.0002 days in our target during their photometric survey 
to find eclipsing $\delta$ Sct stars. As displayed in the third panel of Figure 1, even our spot model did not fully explain 
the TESS light curve of X Tri with exceptional precision. It is not easy to extract reliable pulsating signals from the `W'-shape 
residuals. For useful multifrequency analyses, we made 1000 normal points phase-binning at intervals of 0.001. The mean light curve 
and the residuals from it are presented as the cyan dots in the top and bottom panels of Figure 1. The corresponding residuals 
distributed in BJD are given in Figure 7. We recognized that the binary signals were satisfactorily removed from the observed data 
and the primary eclipse depth varied with time. 

In order to investigate the oscillating features of X Tri, the PERIOD04 software by Lenz \& Breger (2005) was introduced into 
the out-of-eclipse light curve residuals. The amplitude spectrum of the program target was calculated to the Nyquist frequency limit 
of $f_{\rm Nyq} \simeq$ 360 day$^{-1}$. Using the pre-whitening technique (Lee et al. 2014), we found 16 significant signals with SNR 
larger than about 4 for each peak (Breger et al. 1993). The resultant frequency parameters are given in Table 6 and their uncertainties 
were obtained following Kallinger et al. (2008). The synthetic curve prepared from them appears as a solid line in Figure 7, and 
the periodogram of X Tri are presented in Figure 8. There are no conspicuous signals between 60 day$^{-1}$ and $f_{\rm Nyq}$. 

As illustrated in Table 6 and Figure 8, the main signals of X Tri are distributed in the two frequency domains of $<$ 3 day$^{-1}$ 
and 36$-$52 day$^{-1}$. Of our 16 frequencies, the highest amplitude signal $f_2$ corresponds to a frequency of 45.45 $\pm$ 0.41 day$^{-1}$ 
detected by Turner et al. (2014) within their errors, and $f_{14}$ appears to be the orbital harmonic of $2f_{\rm orb}$. We carefully 
explored combination frequencies within the Rayleigh frequency of $R_{\rm f}$ = 0.042 day$^{-1}$, from which our remarks are given 
in the last column of Table 6. We think the low frequencies below $f_{\rm orb}$ might be aliases or artefacts due to systematic trends. 
In contrast, most high frequencies may be identified as possible pulsation signals arising from the hotter and more massive primary.

\section{DISCUSSION AND CONCLUSIONS}

We have analyzed new spectroscopic observations of the semi-detached Algol X Tri with archival TESS data. The RV curves of 
the eclipsing pair were obtained by using the double Gaussian as a fitting function, and we measured the first RVs for 
the cool secondary star. The minimum $\chi^{2}$ method was applied to yield a rotational rate of $v_1$$\sin$$i=84\pm6$ km s$^{-1}$ and 
a surface temperature of $T_{\rm eff,1}=7900 \pm 110$ K for the primary star. With the atmospheric parameters, the TESS light and 
our RV curves let us characterize each binary component. From their simultaneous modeling, we determined the accurate masses and 
radii with less than 1\% uncertainties, as follows: $M_1 = 2.137 \pm 0.018$ $M_\odot$, $M_2 = 1.101 \pm 0.010$ $M_\odot$, 
$R_1 = 1.664 \pm 0.010$ $R_\odot$, and $R_2 = 1.972 \pm 0.010$ $R_\odot$. For X Tri A, the synchronous rotation of $v_{\rm 1,sync}$ = 
86.7 $\pm$ 0.5 km s$^{-1}$ is well matched with our measurement of $v_1$$\sin$$i$, unlike Struve's (1946). At present, our program 
target is expected to be in a state of synchronous rotation. 
Meanwhile, X Tri is a semidetached Algol in which the less massive secondary fills the inner critical Roche lobe, 
while the more massive primary is detached and its fill-out factor is about 0.69. Thus, our target X Tri is a regular Algol, not 
a reverse Algol, i.e. the massive primary fills its Roche lobe. 

Our binary synthesis revealed a third light of $l_3$ = 0.0374, which could originate from either circumbinary objects, as suggested 
by eclipse timing analyses, or other stars near X Tri because each pixel size of the TESS CCD camera is 21 arcsec on the sky 
(Ricker et al. 2015). There is a neighboring star TIC 28391715 (2MASS J02003411+2753233; Gaia EDR3 298268986033892480; 
$T_{\rm p}$ = $+$12.007) with a separation of about 6.5 arcsec from our target star and a magnitude difference of 3.442 mag between 
them in the TESS passband. The brightness difference indicates that TIC 28391715 is about 24 times fainter than X Tri and hence 
contributes within 4 \% of the total light. Thus, most of the $l_3$ source may be the nearby star TIC 28391715. 
The distance to TIC 28391715 is 207.6 $\pm$ 0.7 pc from GAIA EDR3 parallax of 4.816 $\pm$ 0.016 mas (Gaia Collaboration et al. 2021) 
and this is a good match to the Gaia distance 210 $\pm$ 2 pc of X Tri. Further, the TESS targets seem to share a common proper motion; 
$\mu _\alpha \cos \delta$ = 25.893 $\pm$ 0.043 mas yr$^{-1}$ and $\mu _ \delta$ = $-$14.053 $\pm$ 0.046 mas yr$^{-1}$ for X Tri, and 
$\mu _\alpha \cos \delta$ = 26.072 $\pm$ 0.018 mas yr$^{-1}$ and $\mu _ \delta$ = $-$12.953 $\pm$ 0.018 mas yr$^{-1}$ for TIC 28391715. 
So then, TIC 28391715 may be gravitationally bound to the X Tri system, and the separation between them is estimated to be $\sim$1360 AU. 

To extract more reliable oscillating frequencies, we formed a normal light curve at 0.001 phase intervals and the outside-eclipse 
residuals from it were applied to the PERIOD04 software package. Sixteen significant frequencies with SNR $\ga$ 4 were detected 
from the multifrequency analysis, of which the low frequencies in the gravity ($g$)-mode domain are thought to be aliases or 
artefacts. Except for possible combinations within $R_{\rm f}$, our target star X Tri pulsates in the pressure ($p$)-mode frequencies 
between 37 day$^{-1}$ and 48 day$^{-1}$ that originate from the primary star. The pulsation periods of the high frequencies are 
in the range of $P_{\rm pul}$ = 0.021$-$0.027 days. Applying the Table 5 parameters to the relation of 
$\log Q = \log P_{\rm pul} + 0.5 \log g_1 + 0.1M_{\rm bol, 1} + \log T_{\rm eff, 1} - 6.454$ (Petersen \& J\o rgensen 1972), 
we obtained the pulsation constants of $Q$ = 0.014$-$0.018 days. The observed $P_{\rm pul}$ and $Q$ values are typical of 
$\delta$ Sct oscillations (Breger 2000; Aerts et al. 2010). Thus, the detached primary component of X Tri is an oEA pulsator. 
Our pulsation data corresponding to the $p$ modes of $\delta$ Sct variables help to probe the stellar envelope of X Tri A through 
detailed asteroseismic modeling. However, it is difficult to probe the deep stellar interiors near the core region and to determine 
if there is a phase-dependence in these frequencies, because they did not detect in $g$-mode pulsations and were observed in only 
a single TESS band. 
 
Recently, Chen et al. (2022) found that three short-period binaries whose pulsations were discovered by TESS considerably exceed 
the upper limit of $P_{\rm pul}/P_{\rm orb}$ = 0.09$\pm$0.02, meaning that there is no correlation between them in these stars. 
X Tri has a relatively shorter orbital period than other normal oEA binaries (Mkrtichian et al. 2018). 
Its pulsation and orbital period ratios are $P_{\rm pul}/P_{\rm orb} < 0.008$. The physical parameters of our target star 
follow well the empirical relations of the $\log P_{\rm orb} - \log P_{\rm pul}$, $\log P_{\rm pul} - \log (F/M_1)$, and 
$\log P_{\rm pul} - \log g_1$ for the other oEA star with similiar orbital periods, where $F/M_1$ denotes the gravitational pulls 
onto the pulsating components exerted by companions (Soydugan et al. 2006; Liakos 2017; Liakos \& Niarchos 2017). Also, our target 
exhibit almost the same pulsating characteristics as single $\delta$ Sct pulsators.

\acknowledgments{ }
The authors would like to thank the BOAO staffs for assistance during our BOES observations. We also thank the anonymous referee 
for the careful reading and valuable comments. This paper includes data collected by the TESS mission and obtained from MAST. 
Funding for the TESS mission is provided by the NASA Explorer Program. IRAF is distributed by the National Optical Astronomy 
Observatory, which is operated by the Association of Universities for Research in Astronomy (AURA) under a cooperative agreement 
with the National Science Foundation. This research has made use of the Simbad database maintained at CDS, Strasbourg, France, 
and was supported by the KASI grant 2022-1-830-04. K.H. was supported by the grants 2020R1A4A2002885 and 2022R1I1A1A01053320 from 
the National Research Foundation (NRF) of Korea. The research of M.W. was partially supported by the project Cooperatio - Physics 
of the Charles University in Prague.

\clearpage
\begin{figure}
\includegraphics[]{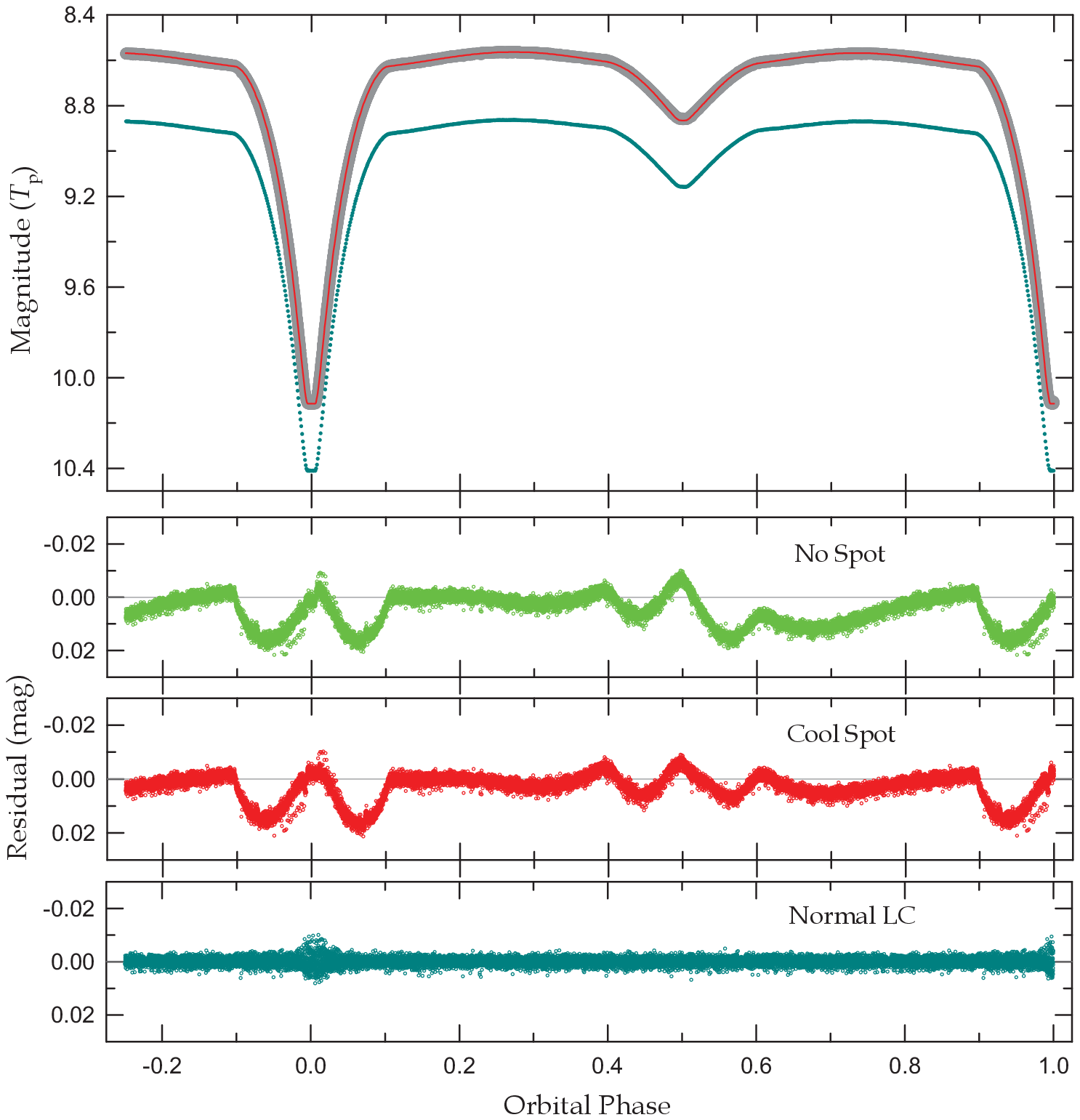}
\caption{Light curve of X Tri with the fitted model. In the top panel, the gray circles are individual measurements from TESS, 
and the red solid curve is computed with a cool spot on the secondary star. The light curve residuals corresponding to the unspotted and 
cool-spot models are plotted in the second and third panels, respectively. The normal points phase-binning at intervals of 0.001 are 
plotted as the cyan dots in the top panel and displaced vertically for clarity. The residuals from the mean curve are shown in the bottom panel. }
\label{Fig1}
\end{figure}

\begin{figure}
\includegraphics[]{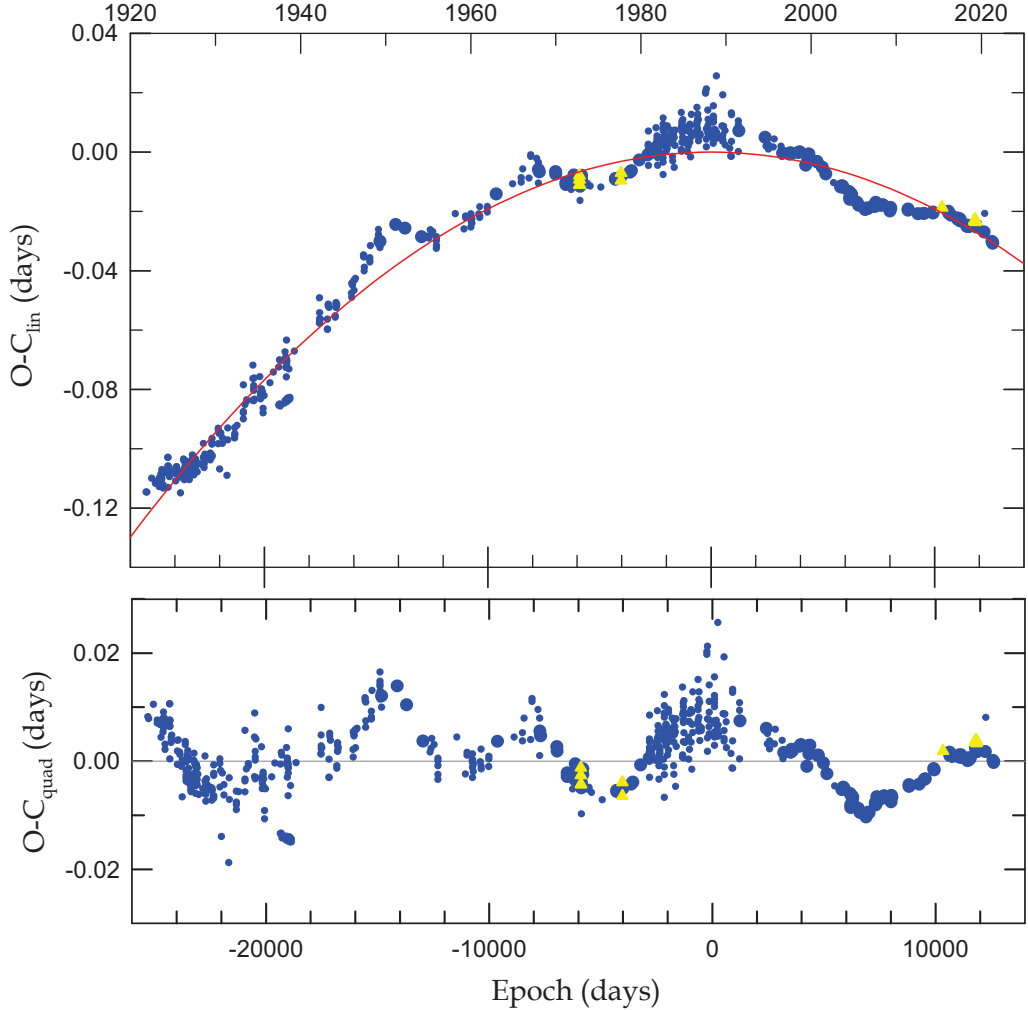}
\caption{Historical $O-C$ diagram of X Tri since discovery, covering over 100 years of observations. The individual primary minima 
are denoted by blue circles and the secondary by yellow triangles. Large symbols correspond to more precise photoelectric or 
CCD measurements. Our new TESS minima are plotted as a bulk of points near the epoch 11800. In the upper panel constructed with 
the linear terms of equation (2), the quadratic fit is given by the red parabolic curve. The timing residuals after subtracting 
the parabolic trend are plotted in the lower panel. The quasi-sinusoidal variations of $O-C$ values up to 0.01 days are clearly visible. }
\label{Fig2}
\end{figure}

\begin{figure}
\includegraphics{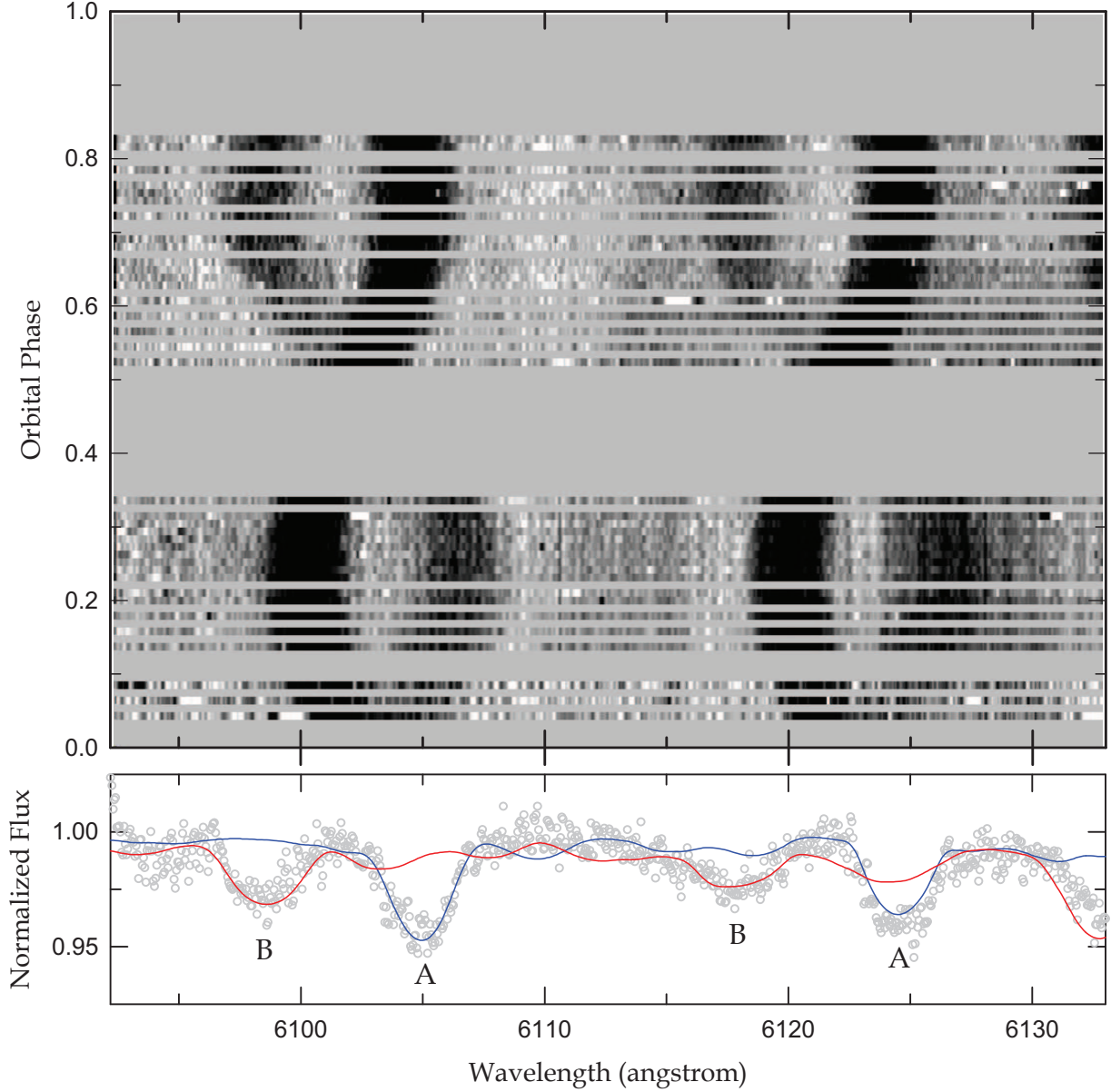}
\caption{The upper panel displays the trailed spectra of X Tri in \ion{Ca}{1} doublet 6103/6122 \AA. In the lower panel, the gray 
circles and blue and red lines represent the observed spectrum at an orbital phase of 0.73 and the synthetic spectra of the primary 
(X Tri A; $T_{\rm eff,1}$ = 7900 K, $\log$ $g_1$ = 4.3, $v_1$$\sin$$i$ = 84 km s$^{-1}$) and secondary (X Tri B; $T_{\rm eff,2}$ = 4990 K, 
$\log$ $g_2$ = 3.9, $v_2$$\sin$$i$ = 103 km s$^{-1}$) components, respectively. }
\label{Fig3}
\end{figure}

\begin{figure}
\includegraphics[]{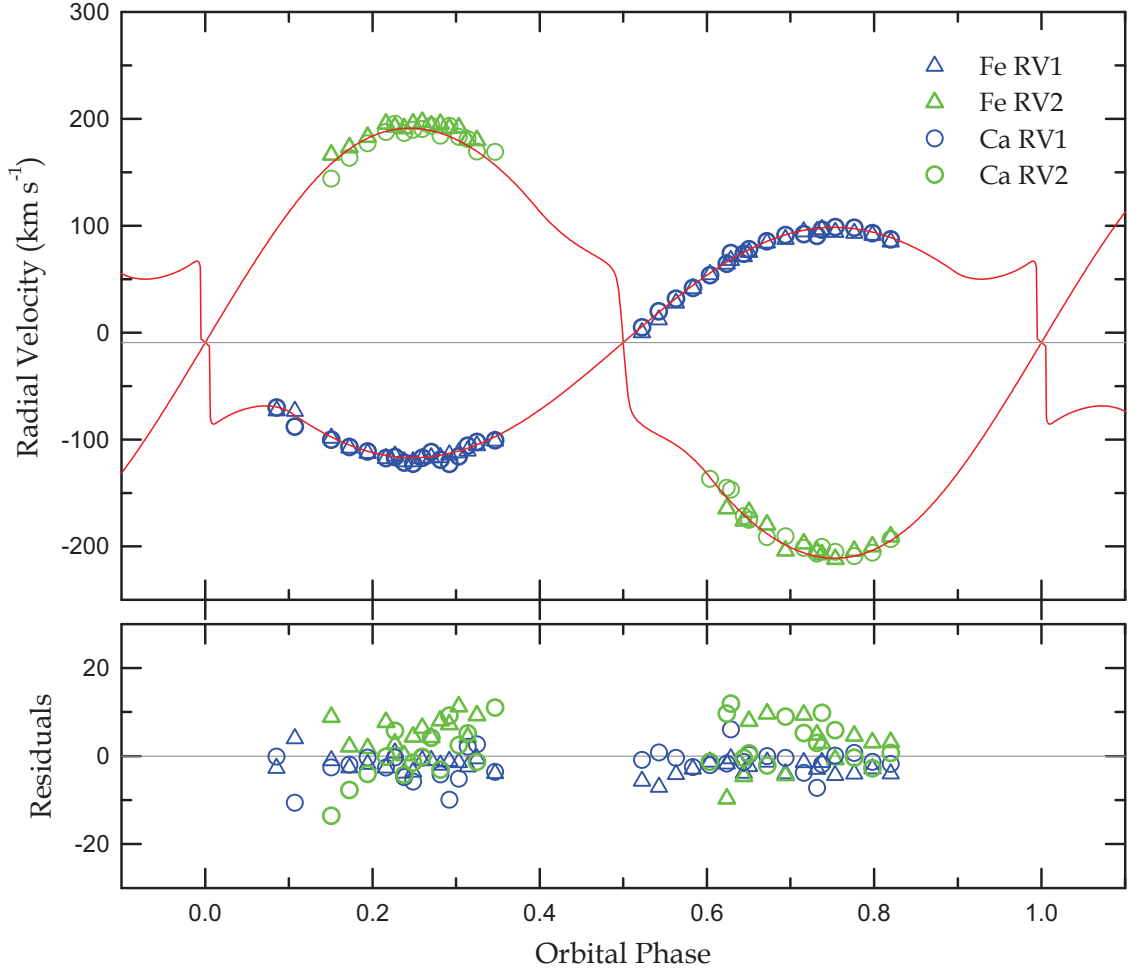}
\caption{RV curves of X Tri with fitted models. The blue and green symbols are the primary and secondary measurements, respectively. 
The red solid curves represent the results from a consistent light and RV curve analysis with the W-D code. The gray line in the upper panel 
represents the system velocity of $-$9.21 km s$^{-1}$. The lower panel shows the RV residuals between observations and models. }
\label{Fig4}
\end{figure}

\begin{figure}
\includegraphics{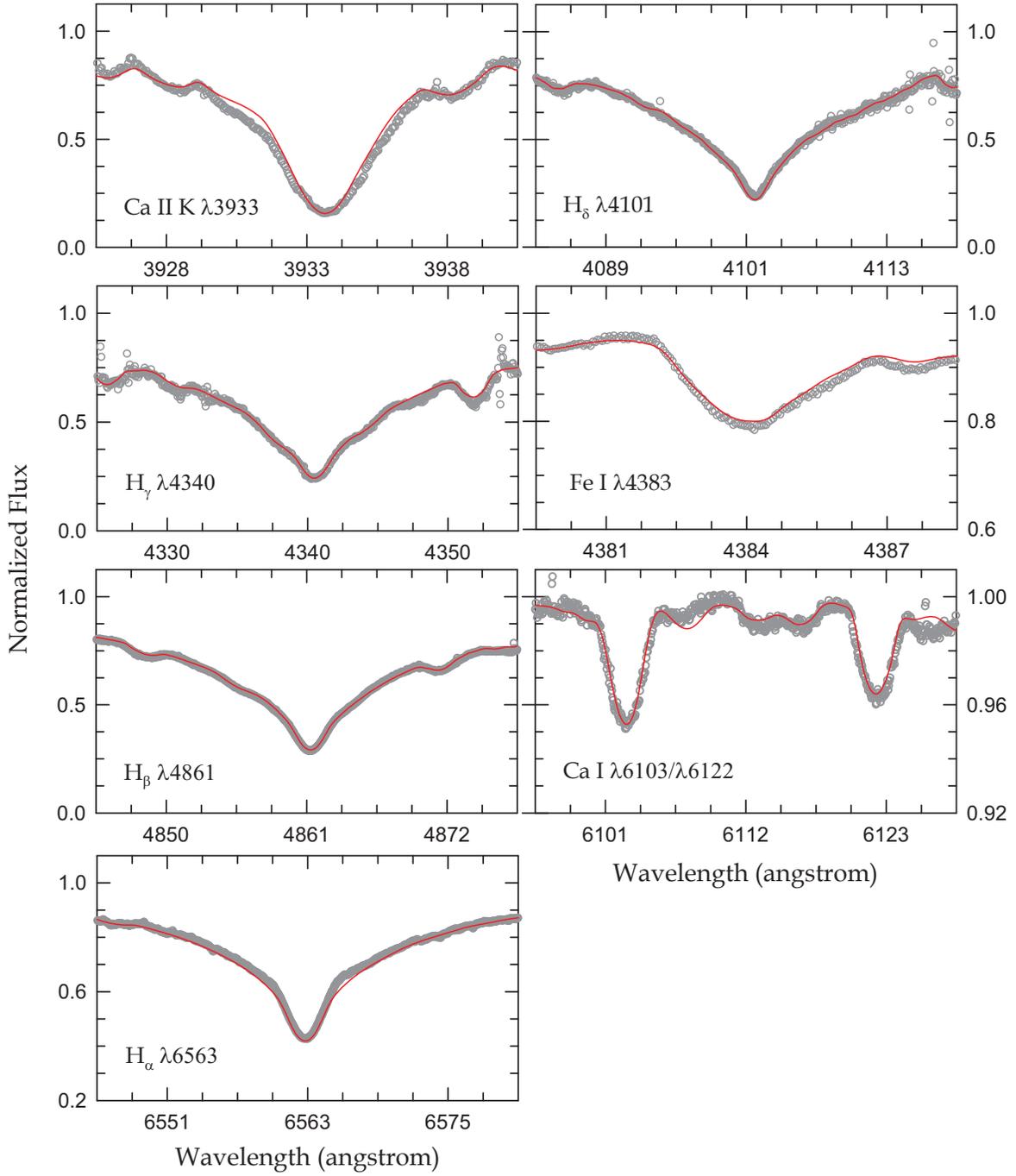}
\caption{Seven spectral regions of the primary star. The gray circles represent the disentangling spectrum obtained by 
the FDB\textsc{inary} code. The red lines denote the synthetic spectra of our best-fit parameters $T_{\rm eff,1}$ = 7900 K and 
$v_1$$\sin$$i$ = 84 km s$^{-1}$. }
\label{Fig5}
\end{figure}

\begin{figure}
\includegraphics[]{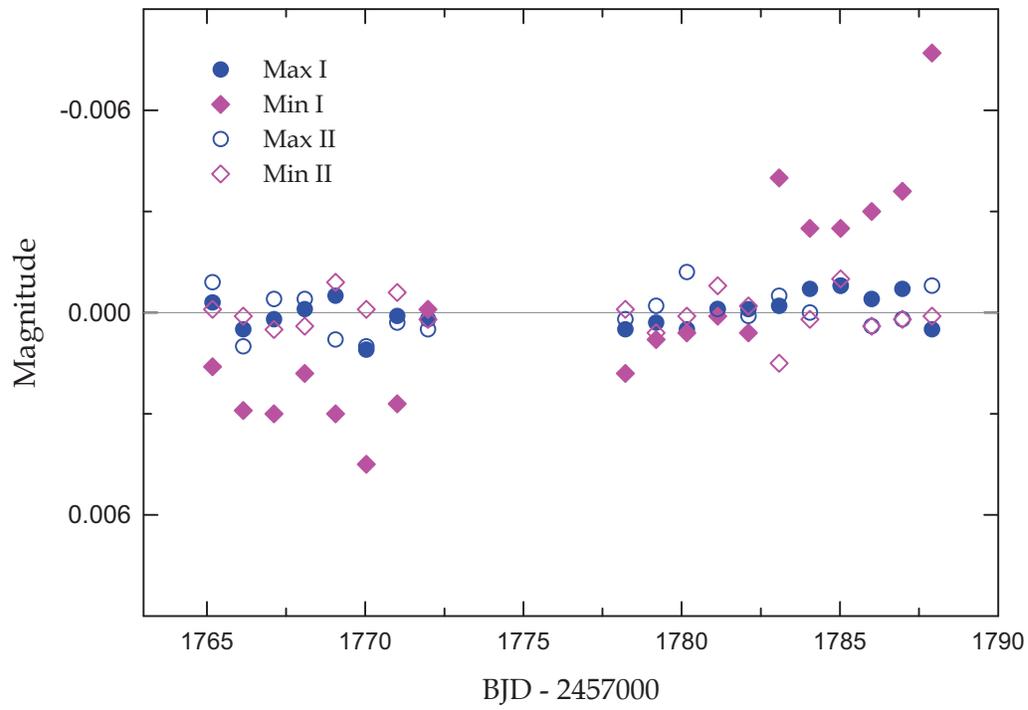}
\caption{Variations in light levels for X Tri at four characteristic phases: Max I, Min I, Max II and Min II. See text in detail. }
 \label{Fig6}
\end{figure}

\begin{figure}
\includegraphics[scale=0.9]{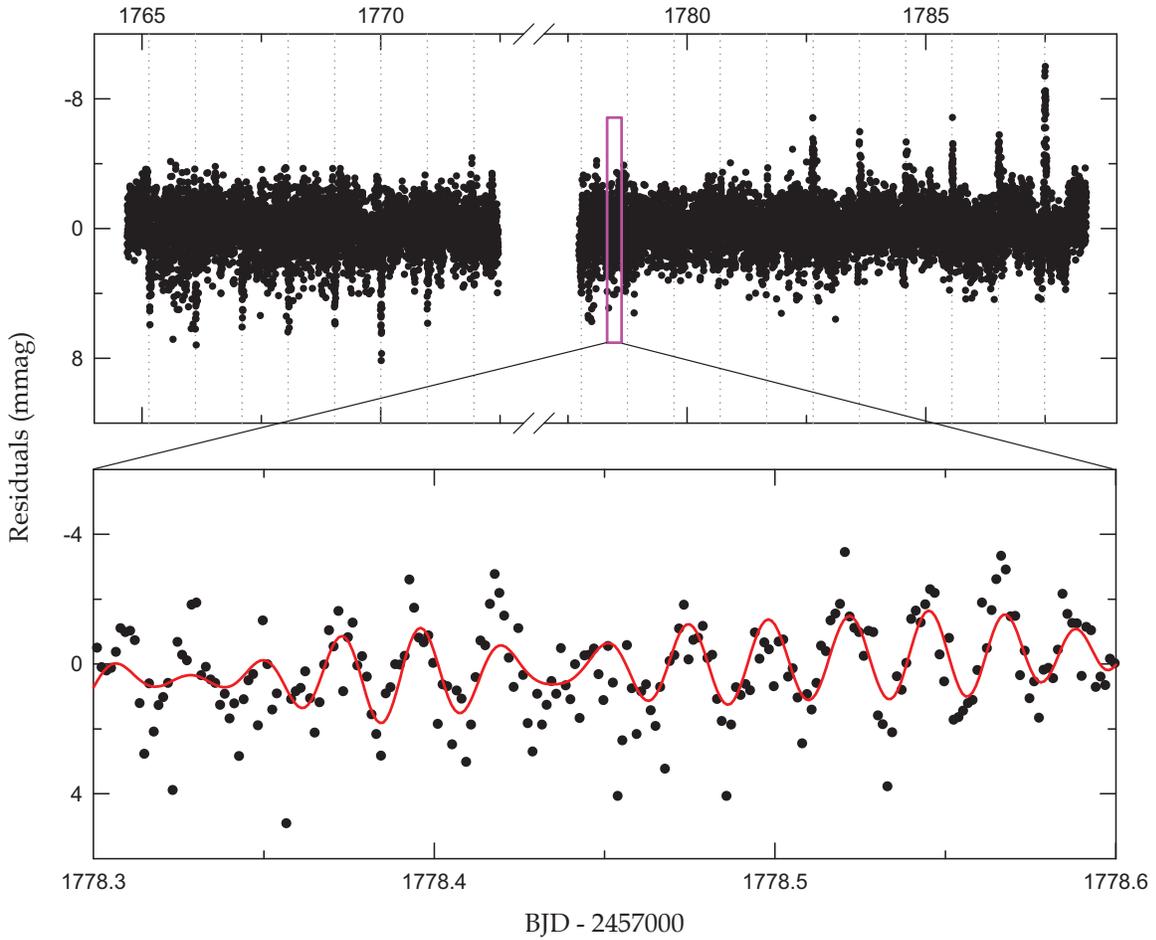}
\caption{Light curve residuals from 1000 normal points distributed in BJD. The vertical dotted lines indicate the primary minimum epochs. 
The lower panel presents a short section of the residuals marked using the inset box in the upper panel. The synthetic curve is computed 
from the 16-frequency fit to the out-of-eclipse part of the residuals. }
\label{Fig7}
\end{figure}

\begin{figure}
\includegraphics[]{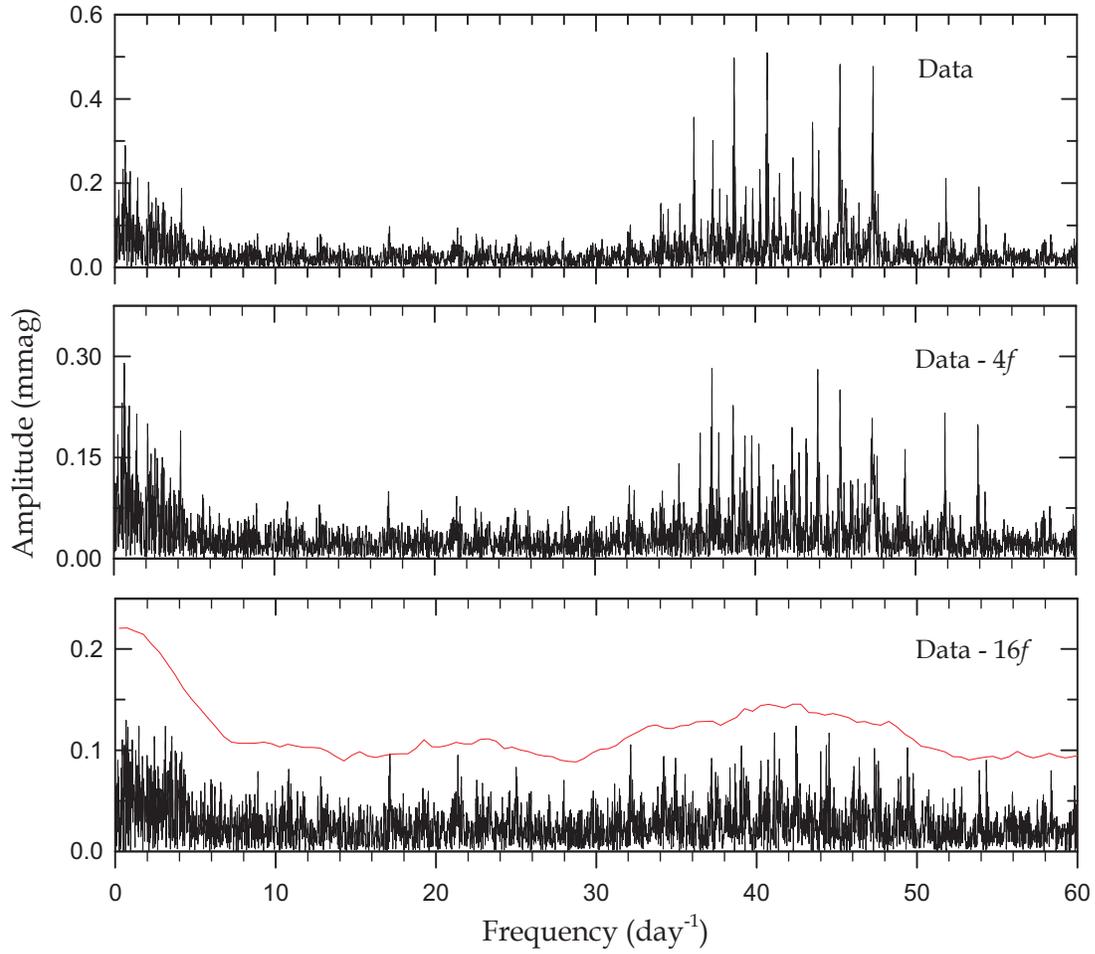}
\caption{Amplitude spectra before (top panel) and after pre-whitening of the first 4 frequencies (middle panel) and all 16 frequencies 
(bottom panel) from the PERIOD04 program. The red line in the bottom panel corresponds to four times the noise spectrum. }
\label{Fig8}
\end{figure}

\clearpage 
\begin{deluxetable}{lccclcc}
\tablewidth{0pt}
\tablecaption{TESS Eclipse Timings for X Tri }
\tablehead{
\colhead{BJD}    & \colhead{Error} & \colhead{Min} & \colhead{BJD}    & \colhead{Error} & \colhead{Min} 
}
\startdata
2,458,765.14863  & $\pm$0.00001    & I             & 2,458,779.23707  & $\pm$0.00010    & II           \\
2,458,765.63585  & $\pm$0.00011    & II            & 2,458,779.72151  & $\pm$0.00001    & I            \\
2,458,766.12013  & $\pm$0.00001    & I             & 2,458,780.20852  & $\pm$0.00010    & II           \\
2,458,766.60747  & $\pm$0.00013    & II            & 2,458,780.69301  & $\pm$0.00001    & I            \\
2,458,767.09168  & $\pm$0.00001    & I             & 2,458,781.18010  & $\pm$0.00009    & II           \\
2,458,767.57895  & $\pm$0.00014    & II            & 2,458,781.66456  & $\pm$0.00001    & I            \\
2,458,768.06322  & $\pm$0.00001    & I             & 2,458,782.15148  & $\pm$0.00010    & II           \\
2,458,768.55024  & $\pm$0.00009    & II            & 2,458,782.63613  & $\pm$0.00001    & I            \\
2,458,769.03474  & $\pm$0.00001    & I             & 2,458,783.12299  & $\pm$0.00009    & II           \\
2,458,769.52184  & $\pm$0.00012    & II            & 2,458,783.60761  & $\pm$0.00001    & I            \\
2,458,770.00627  & $\pm$0.00001    & I             & 2,458,784.09455  & $\pm$0.00010    & II           \\
2,458,770.49336  & $\pm$0.00011    & II            & 2,458,784.57913  & $\pm$0.00001    & I            \\
2,458,770.97779  & $\pm$0.00001    & I             & 2,458,785.06601  & $\pm$0.00010    & II           \\
2,458,771.46491  & $\pm$0.00011    & II            & 2,458,785.55067  & $\pm$0.00001    & I            \\
2,458,771.94933  & $\pm$0.00001    & I             & 2,458,786.03754  & $\pm$0.00009    & II           \\
2,458,772.43605  & $\pm$0.00005    & II            & 2,458,786.52221  & $\pm$0.00001    & I            \\
2,458,777.77844  & $\pm$0.00001    & I             & 2,458,787.00902  & $\pm$0.00008    & II           \\
2,458,778.26567  & $\pm$0.00013    & II            & 2,458,787.49374  & $\pm$0.00001    & I            \\
2,458,778.74997  & $\pm$0.00001    & I             & 2,458,787.98039  & $\pm$0.00007    & II           \\
\enddata
\end{deluxetable}

\begin{deluxetable}{lrrrrrrrrrr}                                                                                            
\tablewidth{0pt}                    
\tabletypesize{\small}   
\rotate                                                                          
\tablecaption{Radial velocities of X Tri }                                                                            
\tablehead{ 
                       &                   & \multicolumn{4}{c}{\ion{Fe}{1} $\lambda$4957}                                                         && \multicolumn{4}{c}{\ion{Ca}{1} $\lambda$6103}                                                         \\ [1.0mm] \cline{3-6} \cline{8-11} \\ [-2.0ex]
\colhead{BJD}          & \colhead{UT date} & \colhead{$V_{1}$}       & \colhead{$\sigma_1$}    & \colhead{$V_{2}$}       & \colhead{$\sigma_2$}    && \colhead{$V_{1}$}       & \colhead{$\sigma_1$}    & \colhead{$V_{2}$}       & \colhead{$\sigma_2$}    \\ 
\colhead{(2,457,000+)} &                   & \colhead{(km s$^{-1}$)} & \colhead{(km s$^{-1}$)} & \colhead{(km s$^{-1}$)} & \colhead{(km s$^{-1}$)} && \colhead{(km s$^{-1}$)} & \colhead{(km s$^{-1}$)} & \colhead{(km s$^{-1}$)} & \colhead{(km s$^{-1}$)} 
}                                                                                                                    
\startdata                                                                                                                                                                        
1,820.0618             & 2019-12-02        & $   0.1$                & 6.1                     & $      $                &                         && $   4.8$                &  2.2                    & $      $                &                         \\
1,820.0816             & 2019-12-02        & $  12.1$                & 4.4                     & $      $                &                         && $  19.9$                &  6.2                    & $      $                &                         \\
1,820.1013             & 2019-12-02        & $  27.9$                & 4.9                     & $      $                &                         && $  31.6$                &  6.3                    & $      $                &                         \\
1,820.1210             & 2019-12-02        & $  41.4$                & 5.9                     & $      $                &                         && $  41.8$                &  6.7                    & $      $                &                         \\
1,820.1408             & 2019-12-02        & $  54.4$                & 7.6                     & $      $                &                         && $  53.6$                &  5.0                    & $-136.8$                &  8.5                    \\ 
1,820.1605             & 2019-12-02        & $  64.3$                & 5.6                     & $-164.5$                & 10.0                    && $  64.3$                &  7.7                    & $-145.2$                & 10.7                    \\
1,820.1803             & 2019-12-02        & $  71.3$                & 4.5                     & $-176.0$                &  8.2                    && $  73.8$                &  4.1                    & $-171.9$                & 12.4                    \\
2,145.1004             & 2020-10-22        & $ -72.8$                & 8.8                     & $      $                &                         && $ -70.2$                &  6.7                    & $      $                &                         \\
2,145.1215             & 2020-10-22        & $ -73.4$                & 7.4                     & $      $                &                         && $ -88.0$                & 11.9                    & $      $                &                         \\
2,145.2379             & 2020-10-22        & $-115.3$                & 6.4                     & $ 192.6$                &  9.0                    && $-116.4$                &  7.0                    & $ 195.5$                &  6.7                    \\ 
2,145.2590             & 2020-10-22        & $-120.6$                & 3.1                     & $ 195.6$                &  7.0                    && $-122.8$                &  7.3                    & $ 189.7$                &  9.0                    \\ 
2,145.2801             & 2020-10-22        & $-116.9$                & 4.5                     & $ 193.7$                &  4.5                    && $-111.8$                &  5.3                    & $ 193.5$                &  5.5                    \\ 
2,145.3013             & 2020-10-22        & $-114.1$                & 5.4                     & $ 191.5$                &  5.5                    && $-122.9$                &  6.8                    & $ 193.5$                &  9.1                    \\ 
2,145.3224             & 2020-10-22        & $-110.5$                & 6.8                     & $ 180.4$                &  8.2                    && $-106.0$                &  4.3                    & $ 181.2$                &  5.3                    \\ 
2,146.1355             & 2020-10-23        & $ -98.6$                & 3.1                     & $ 166.4$                &  6.1                    && $-100.2$                &  9.7                    & $ 143.9$                & 13.5                    \\
2,146.1567             & 2020-10-23        & $-107.8$                & 6.4                     & $ 173.3$                &  6.8                    && $-107.2$                &  5.4                    & $ 163.4$                & 10.2                    \\
2,146.1778             & 2020-10-23        & $-112.8$                & 3.9                     & $ 183.0$                &  4.1                    && $-111.3$                &  5.2                    & $ 177.1$                & 10.3                    \\
2,146.1989             & 2020-10-23        & $-117.5$                & 5.4                     & $ 195.5$                &  9.6                    && $-117.5$                &  4.5                    & $ 187.6$                &  8.5                    \\ 
2,146.2201             & 2020-10-23        & $-120.6$                & 5.7                     & $ 191.1$                &  8.4                    && $-121.6$                &  6.3                    & $ 186.7$                &  9.6                    \\ 
2,146.2412             & 2020-10-23        & $-117.7$                & 5.0                     & $ 197.1$                &  5.5                    && $-117.1$                &  4.2                    & $ 190.5$                &  7.2                    \\ 
2,146.2623             & 2020-10-23        & $-116.7$                & 2.8                     & $ 195.3$                &  4.5                    && $-118.9$                &  4.0                    & $ 184.1$                & 10.0                    \\
2,146.2835             & 2020-10-23        & $-112.2$                & 5.0                     & $ 191.8$                &  5.1                    && $-115.9$                &  5.2                    & $ 183.0$                &  7.6                    \\ 
2,146.3046             & 2020-10-23        & $-105.5$                & 4.0                     & $ 180.0$                &  5.9                    && $-102.3$                &  9.7                    & $ 169.4$                & 17.6                    \\
2,146.3257             & 2020-10-23        & $-101.3$                & 2.6                     & $      $                &                         && $-101.0$                &  5.6                    & $ 168.9$                &  7.8                    \\ 
2,167.0020             & 2020-11-13        & $  67.7$                & 3.4                     & $      $                &                         && $  74.3$                &  5.6                    & $-147.0$                & 16.9                    \\
2,167.0230             & 2020-11-13        & $  75.3$                & 3.0                     & $-167.8$                &  9.8                    && $  77.9$                &  5.2                    & $-175.2$                & 17.1                    \\
2,167.0441             & 2020-11-13        & $  84.1$                & 5.2                     & $-179.7$                &  5.0                    && $  85.4$                &  3.7                    & $-191.5$                &  9.8                    \\ 
2,167.0654             & 2020-11-13        & $  87.6$                & 5.4                     & $-204.1$                &  6.6                    && $  91.1$                &  6.0                    & $-190.8$                &  9.1                    \\ 
2,167.0868             & 2020-11-13        & $  94.4$                & 5.5                     & $-197.5$                &  9.0                    && $  92.0$                &  4.9                    & $-201.7$                & 12.3                    \\
2,167.1078             & 2020-11-13        & $  96.7$                & 4.5                     & $-207.7$                &  7.3                    && $  96.2$                &  6.7                    & $-200.8$                &  8.7                    \\ 
2,169.0452             & 2020-11-15        & $  94.7$                & 4.2                     & $-205.1$                &  8.0                    && $  90.4$                &  6.0                    & $-207.0$                & 22.0                    \\
2,169.0664             & 2020-11-15        & $  94.2$                & 5.0                     & $-212.0$                &  7.8                    && $  98.6$                &  6.5                    & $-205.4$                &  9.6                    \\ 
2,169.0883             & 2020-11-15        & $  93.3$                & 5.7                     & $-204.4$                &  6.7                    && $  98.0$                &  5.7                    & $-209.4$                &  8.0                    \\ 
2,169.1095             & 2020-11-15        & $  91.3$                & 3.0                     & $-200.2$                &  8.2                    && $  92.8$                &  7.1                    & $-206.1$                &  8.2                    \\ 
2,169.1308             & 2020-11-15        & $  85.0$                & 4.4                     & $-190.9$                &  7.7                    && $  87.2$                &  5.3                    & $-193.4$                & 10.1                    \\
\enddata 
\end{deluxetable}

\begin{deluxetable}{lcccc}
\tablewidth{0pt}
\tablecaption{Light levels of X Tri at four different phases }
\tablehead{
\colhead{Mean Time}  & \colhead{Min I}           & \colhead{Max I}            & \colhead{Min II}           & \colhead{Max II}         \\ 
\colhead{(BJD)}      & \colhead{(mag)}           & \colhead{(mag)}            & \colhead{(mag)}            & \colhead{(mag)}              
}
\startdata
2,458,765.17500      & 10.0972$\pm$0.0252        & 8.5652$\pm$0.0013          & 8.8562$\pm$0.0045          & 8.5708$\pm$0.0014        \\
2,458,766.14726      & 10.0985$\pm$0.0275        & 8.5660$\pm$0.0014          & 8.8564$\pm$0.0040          & 8.5727$\pm$0.0020        \\
2,458,767.11951      & 10.0986$\pm$0.0258        & 8.5657$\pm$0.0019          & 8.8568$\pm$0.0046          & 8.5713$\pm$0.0014        \\
2,458,768.09176      & 10.0974$\pm$0.0263        & 8.5654$\pm$0.0024          & 8.8567$\pm$0.0039          & 8.5713$\pm$0.0013        \\
2,458,769.06401      & 10.0986$\pm$0.0254        & 8.5650$\pm$0.0021          & 8.8554$\pm$0.0033          & 8.5725$\pm$0.0017        \\
2,458,770.03626      & 10.1001$\pm$0.0268        & 8.5666$\pm$0.0011          & 8.8562$\pm$0.0036          & 8.5727$\pm$0.0011        \\
2,458,771.00851      & 10.0983$\pm$0.0252        & 8.5656$\pm$0.0024          & 8.8557$\pm$0.0040          & 8.5720$\pm$0.0008        \\
2,458,771.98006      & 10.0955$\pm$0.0265        & 8.5657$\pm$0.0019          & 8.8565$\pm$0.0034          & 8.5722$\pm$0.0008        \\
2,458,778.22116      & 10.0974$\pm$0.0255        & 8.5660$\pm$0.0013          & 8.8562$\pm$0.0044          & 8.5719$\pm$0.0008        \\
2,458,779.19340      & 10.0964$\pm$0.0254        & 8.5658$\pm$0.0006          & 8.8569$\pm$0.0042          & 8.5715$\pm$0.0016        \\
2,458,780.16563      & 10.0962$\pm$0.0256        & 8.5660$\pm$0.0013          & 8.8564$\pm$0.0037          & 8.5705$\pm$0.0013        \\
2,458,781.13787      & 10.0957$\pm$0.0262        & 8.5654$\pm$0.0010          & 8.8555$\pm$0.0026          & 8.5716$\pm$0.0017        \\
2,458,782.11010      & 10.0962$\pm$0.0248        & 8.5654$\pm$0.0012          & 8.8561$\pm$0.0037          & 8.5718$\pm$0.0011        \\
2,458,783.08232      & 10.0916$\pm$0.0260        & 8.5653$\pm$0.0012          & 8.8578$\pm$0.0033          & 8.5712$\pm$0.0010        \\
2,458,784.05455      & 10.0931$\pm$0.0254        & 8.5648$\pm$0.0019          & 8.8565$\pm$0.0035          & 8.5717$\pm$0.0005        \\
2,458,785.02678      & 10.0931$\pm$0.0257        & 8.5647$\pm$0.0010          & 8.8553$\pm$0.0025          & 8.5709$\pm$0.0008        \\
2,458,785.99900      & 10.0926$\pm$0.0249        & 8.5651$\pm$0.0013          & 8.8567$\pm$0.0043          & 8.5721$\pm$0.0012        \\
2,458,786.97122      & 10.0920$\pm$0.0258        & 8.5648$\pm$0.0007          & 8.8565$\pm$0.0042          & 8.5719$\pm$0.0009        \\
2,458,787.90941      & 10.0879$\pm$0.0254        & 8.5660$\pm$0.0015          & 8.8564$\pm$0.0043          & 8.5709$\pm$0.0011        \\ [1.0mm] \cline{1-5} \\ [-2.0ex]
Average              & 10.0956$\pm$0.0031        & 8.5655$\pm$0.0005          & 8.8563$\pm$0.0006          & 8.5717$\pm$0.0006        \\
\enddata
\end{deluxetable}

\begin{deluxetable}{lcc}
\tablewidth{0pt} 
\tablecaption{Light and RV Parameters of X Tri }
\tablehead{
\colhead{Parameter}                & \colhead{Primary}  & \colhead{Secondary}                                                  
}                                                                                                                                     
\startdata                                                                                                                            
$T_0$ (BJD)                        & \multicolumn{2}{c}{2,458,765.148597$\pm$0.000028}     \\
$P_{\rm orb}$ (day)                & \multicolumn{2}{c}{0.9715319$\pm$0.0000018}           \\
$a$ ($R_\odot$)                    & \multicolumn{2}{c}{6.105$\pm$0.021}                   \\
$\gamma$ (km s$^{-1}$)             & \multicolumn{2}{c}{$-$9.21$\pm$0.29}                  \\
$K_1$ (km s$^{-1}$)                & \multicolumn{2}{c}{108.15$\pm$0.50}                   \\
$K_2$ (km s$^{-1}$)                & \multicolumn{2}{c}{209.83$\pm$0.98}                   \\
$q$                                & \multicolumn{2}{c}{0.5154$\pm$0.0035}                 \\
$i$ (deg)                          & \multicolumn{2}{c}{88.90$\pm$0.14}                    \\
$T_{\rm eff}$ (K)                  & 7900$\pm$110           & 4990$\pm$50                  \\
$\Omega$                           & 4.220$\pm$0.014        & 2.905                        \\
$X_{\rm bol}$, $Y_{\rm bol}$       & 0.673, 0.203           & 0.642, 0.165                 \\
$x_{T_{\rm P}}$, $y_{T_{\rm P}}$   & 0.574$\pm$0.031, 0.214 & 0.695$\pm$0.023, 0.188       \\
$l$/($l_{1}$+$l_{2}$+$l_{3}$)      & 0.7210$\pm$0.0031      & 0.2416                       \\
$l_{3}$$\rm ^a$                    & \multicolumn{2}{c}{0.0374$\pm$0.0024}                 \\
$r$ (pole)                         & 0.2687$\pm$0.0013      & 0.3022$\pm$0.0011            \\
$r$ (point)                        & 0.2786$\pm$0.0015      & 0.4323$\pm$0.0013            \\
$r$ (side)                         & 0.2728$\pm$0.0013      & 0.3154$\pm$0.0012            \\
$r$ (back)                         & 0.2766$\pm$0.0014      & 0.3479$\pm$0.0012            \\
$r$ (volume)$\rm ^b$               & 0.2728$\pm$0.0014      & 0.3232$\pm$0.0012            \\ [0.5mm]
\multicolumn{3}{l}{Spot Parameters:}                                                       \\
Colatitude (deg)                   &                        & 59.9$\pm$2.4                 \\        
Longitude (deg)                    &                        & 335.9$\pm$1.9                \\        
Radius (deg)                       &                        & 16.8$\pm$2.8                 \\        
$T$$\rm _{spot}$/$T$$\rm _{local}$ &                        & 0.885$\pm$0.039              \\
\enddata
\tablenotetext{a}{Value at 0.25 orbital phase. }
\tablenotetext{b}{Mean volume radius. }
\end{deluxetable}

\begin{deluxetable}{lcc}
\tablewidth{0pt} 
\tablecaption{Absolute Parameters of X Tri }
\tablehead{
\colhead{Parameter}           & \colhead{Primary}   & \colhead{Secondary}                                                  
}                                                                                                                                     
\startdata                                                                                                                            
$M$ ($M_\odot$)               & 2.137$\pm$0.018     & 1.101$\pm$0.010             \\
$R$ ($R_\odot$)               & 1.664$\pm$0.010     & 1.972$\pm$0.010             \\
$\log$ $g$ (cgs)              & 4.325$\pm$0.006     & 3.890$\pm$0.006             \\
$\rho$ ($\rho_\odot$)         & 0.464$\pm$0.010     & 0.1439$\pm$0.002            \\
$v_{\rm sync}$ (km s$^{-1}$)  & 86.71$\pm$0.54      & 102.72$\pm$0.52             \\
$v$$\sin$$i$ (km s$^{-1}$)    & 84$\pm$6            & \,                          \\
$T_{\rm eff}$ (K)             & 7900$\pm$110        & 4990$\pm$50                 \\
$L$ ($L_\odot$)               & 9.67$\pm$0.55       & 2.16$\pm$0.09               \\
$M_{\rm bol}$ (mag)           & 2.267$\pm$0.062     & 3.894$\pm$0.045             \\
BC (mag)                      & 0.022$\pm$0.001     & $-$0.292$\pm$0.017          \\
$M_{\rm V}$ (mag)             & 2.245$\pm$0.062     & 4.186$\pm$0.048             \\
Distance (pc)                 & \multicolumn{2}{c}{215$\pm$6}                     \\
\enddata
\end{deluxetable}

\begin{deluxetable}{lrccrc}
\tablewidth{0pt}
\tablecaption{Results of the multiple frequency analysis for X Tri$\rm ^a$ }
\tablehead{
             & \colhead{Frequency}    & \colhead{Amplitude} & \colhead{Phase} & \colhead{SNR$\rm ^b$}  & \colhead{Remark}    \\
             & \colhead{(day$^{-1}$)} & \colhead{(mmag)}    & \colhead{(rad)} &                        &
} 
\startdata 
$f_{1}$      & 40.6840$\pm$0.0016     & 0.322$\pm$0.062     & 1.55$\pm$0.57   &   8.86                 &                     \\
$f_{2}$      & 45.2287$\pm$0.0010     & 0.486$\pm$0.057     & 0.14$\pm$0.35   &  14.49                 &                     \\
$f_{3}$      & 36.1139$\pm$0.0014     & 0.320$\pm$0.054     & 1.15$\pm$0.49   &  10.16                 & $2f_1-f_2$          \\
$f_{4}$      & 43.5160$\pm$0.0014     & 0.352$\pm$0.059     & 0.06$\pm$0.49   &  10.19                 &                     \\
$f_{5}$      &  0.6402$\pm$0.0032     & 0.253$\pm$0.095     & 5.56$\pm$1.10   &   4.57                 &                     \\
$f_{6}$      & 37.3009$\pm$0.0017     & 0.269$\pm$0.055     & 2.28$\pm$0.60   &   8.39                 &                     \\
$f_{7}$      & 43.8975$\pm$0.0019     & 0.266$\pm$0.059     & 2.10$\pm$0.64   &   7.78                 &                     \\
$f_{8}$      & 45.2902$\pm$0.0015     & 0.330$\pm$0.058     & 4.25$\pm$0.51   &   9.82                 & $2f_1-f_3$          \\
$f_{9}$      & 38.6236$\pm$0.0015     & 0.321$\pm$0.056     & 2.74$\pm$0.51   &   9.76                 &                     \\
$f_{10}$     & 51.8274$\pm$0.0017     & 0.212$\pm$0.042     & 1.13$\pm$0.58   &   8.60                 & $2f_2-f_9$          \\
$f_{11}$     & 42.2950$\pm$0.0025     & 0.209$\pm$0.063     & 1.11$\pm$0.88   &   5.72                 &                     \\
$f_{12}$     &  0.9539$\pm$0.0041     & 0.193$\pm$0.094     & 4.57$\pm$1.43   &   3.50                 &                     \\
$f_{13}$     & 47.2891$\pm$0.0015     & 0.298$\pm$0.054     & 5.47$\pm$0.53   &   9.43                 &                     \\
$f_{14}$     &  2.0964$\pm$0.0036     & 0.207$\pm$0.088     & 1.62$\pm$1.25   &   4.01                 & $2f_{\rm orb}$      \\
$f_{15}$     &  0.5066$\pm$0.0040     & 0.198$\pm$0.094     & 6.06$\pm$1.39   &   3.61                 &                     \\
$f_{16}$     & 39.7789$\pm$0.0034     & 0.147$\pm$0.059     & 1.17$\pm$1.18   &   4.24                 & $2f_4-f_{13}$       \\

\enddata                                                                                                                           
\tablenotetext{a}{Frequencies are listed in order of detection. }
\tablenotetext{b}{Calculated in a range of 5 day$^{-1}$ around each frequency. }
\end{deluxetable}

\end{document}